\documentclass[usenatbib]{mn2e}
\usepackage{graphicx,epsfig,aas_macros}
\usepackage{graphics}
\usepackage{amsmath}

\begin{document}
\title{Orbital phase resolved spectroscopy of GX 301-2 with MAXI}
\date{}
\author[Nazma Islam and Biswajit Paul]{Nazma Islam$^{1,2}$\thanks{E-mail:
nazma@rri.res.in;} and Biswajit 
Paul$^{1}$ \\
$^{1}$Raman Research Institute, Sadashivnagar, Bangalore-560080, India\\
$^{2}$Joint Astronomy Programme, Indian Institute of Science, Bangalore-560012, India}

\maketitle

\begin{abstract}
GX 301-2, a bright HMXB with an orbital period of 41.5 days, exhibits stable periodic orbital intensity modulations with a strong pre-periastron
X-ray flare. Several models have been proposed to explain the accretion at different orbital phases, invoking accretion via stellar wind, equatorial disk, 
and accretion stream from the companion star. We present results from exhaustive orbital phase resolved spectroscopic measurements of GX 301-2 using data from 
the Gas Slit Camera onboard {\it MAXI}. Using spectroscopic analysis of the {\it MAXI} data with unprecendented orbital coverage for many orbits continuously, 
we have found a strong orbital dependence of the absorption column density and equivalent width of the iron emission line. 
A very large equivalent width of the iron line along with a small value of the column density in the orbital phase range 0.10-0.30 after the periastron passage 
indicates presence of high density absorbing matter behind the neutron star in these orbital phase range. A low energy excess is also found in the spectrum at orbital 
phases around the pre-periastron X-ray flare. The orbital dependence of these parameters are then used to examine the various models about mode of accretion 
onto the neutron star in GX 301-2.
\end{abstract}


\section{Introduction}

GX 301-2 is a High Mass X-ray Binary (HMXB) system with an X-ray pulsar and a B-emission line hypergiant star (B1 Ia+) WRAY 997. The orbital period of the
binary system is  $\sim$ 41.5 days and the neutron star has a large spin period of $\sim$ 685 s \citep{koh1997}. The distance to GX 301-2 is estimated to be 
$\sim$ 3.1 kpc \citep{coleiro2013}. It is a highly eccentric binary system with e=0.462 \citep{sato1986,koh1997} and has a periodically variable X-ray intensity.\\
The orbital characteristics of GX 301-2 have been extensively studied with various X-ray instruments in different energy bands
\citep[{\it CGRO-BATSE} (15-55 keV) - ][]{koh1997} \citep[{\it RXTE ASM} (2-12 keV) - ][]{pravdo2001,leahy2002,leahy2008} 
\citep[{\it SuperAGILE} (20-60 keV) - ][]{evangelista2010}. It has three intensity regimes: bright phase during X-ray flare (pre-periastron passage 
around orbital phase 0.95), dim or low intensity phase (after periastron passage around orbital phase 0.15-0.3) and intermediate intensity phase (during the 
apastron passage around orbital phase 0.5) \citep{koh1997,leahy2002}. A strong X-ray flare occurs before the periastron passage as well as a medium intensity 
peak is observed at the apastron passage, indicating accretion onto the neutron star due to both spherical stellar wind along with a possible equatorial 
disk or accretion stream \citep{sato1986,stevens1988,haberl1991,layton1998,pravdo2001,leahy2008}. \\
The X-ray spectrum of GX 301-2 have been studied in different orbital phases with {\it TENMA} \citep{leahy1990}, {\it ASCA} \citep{saraswat1996,endo2002}, 
{\it RXTE PCA} \citep{mukherjee2004}, {\it Chandra} \citep{watanabe2003}, {\it BeppoSAX} \citep{labarbera2005}, {\it XMM-Newton} \citep{furst2011} and 
{\it Suzaku} \citep{suchy2012}. 
It has a highly absorbed X-ray spectrum with a partial covering high energy cutoff power-law component and several emission lines. A soft excess component 
is found in the X-ray spectrum from {\it EXOSAT} and {\it ASCA} observations \citep{haberl1991,saraswat1996}. GX 301-2 has a very high line of sight photo-electric absorption, which is
attributed to the dense circumstellar environment in which the neutron star moves. The column density varies strongly with orbital phase with certain amount of 
clumpiness attributed to the stellar wind \citep{pravdo2001,mukherjee2004,leahy2008}. A prominent Fe K$\alpha$ line is found to exist in almost all orbital 
phases of GX 301-2. This fluorescence line is produced due to reprocessing of X-ray photons from the pulsar by the circumstellar matter and sometimes 
shows evidence of a Compton recoil (Watanabe et al. 2003). \\
The equivalent width of the Fe K$\alpha$ line depends on the distribution (geometry and column density) of the surrounding matter
\citep{inoue1985,makino1985,makishima1986,leahy1993,kallman2004}.
Therefore, by comparing the equivalent width of Fe K$\alpha$ line with N$_{H}$, we can study the distribution of circumstellar matter around the 
neutron star at different orbital phases. This can be further used in examining the various accretion models of GX 301-2, in which the accretion
rate and the reprocessing environment have strong orbital phase dependence. In this work, we have investigated the accretion phenomena in GX 301-2 using the 
orbital variation of column density of absorbing matter and the relation between the line equivalent width and N$_{H}$ as observational parameters. 
For such studies, it is very useful to have long term orbital light-curves and spectrum measured at different orbital phases with uniform phase coverage. 
The Monitor of All Sky X-ray Image ({\it MAXI}), which has both all sky coverage and moderate energy resolution, is well suited for detailed studies of the 
orbital light-curves and the orbital phase resolved spectrum of bright X-ray sources. \\
In the present work, we have used light curves of GX 301-2 in the 2-20 keV energy range obtained over four years with the {\it MAXI-GSC} to create multi-band
orbital intensity profiles. From spectroscopic analysis of the {\it MAXI} data with unprecendented orbital coverage for many orbits continuously, we have measured 
the orbital dependence of the absorption column density and equivalent width of the Fe K$\alpha$ line. The orbital phase dependence of the column density
and the line equivalent width are then used to examine the various models about the distribution of circumstellar matter and the mode of accretion onto the 
neutron star in GX 301-2. 

\section{Data and Analysis}
\subsection{Monitor of All Sky X-ray image}
{\it MAXI} is an all sky monitor operated on the Kibo module of the International Space Station (ISS) \citep{matsuoka2009}. It has the best sensitivity and
energy resolution amongst all operating all sky monitors. It has two instruments onboard: Solid state Slit Camera (SSC), operating between 0.7-7 keV 
\citep{tomida2011} and Gas Slit Camera (GSC), operating in the energy range 2-20 keV \citep{mihara2011}. We have used archival data from GSC
\footnote{http://maxi.riken.jp/top} in this work. The GSC consists of six units of large area position sensitive Xenon proportional counters with an 
instantaneous field of view of 1.5$^{\circ} \times$ 160$^{\circ}$. It has 85\% coverage of the entire sky in every 92 minutes of the ISS orbit. The typical 
daily exposure of GSC is 1500 cm$^{2}$s. The in-orbit performance of GSC is summarised in \citet{sugizaki2011}.

\subsection{Energy resolved orbital intensity profiles}
\begin{figure*}
\centering
\includegraphics[angle=-90,scale=0.25]{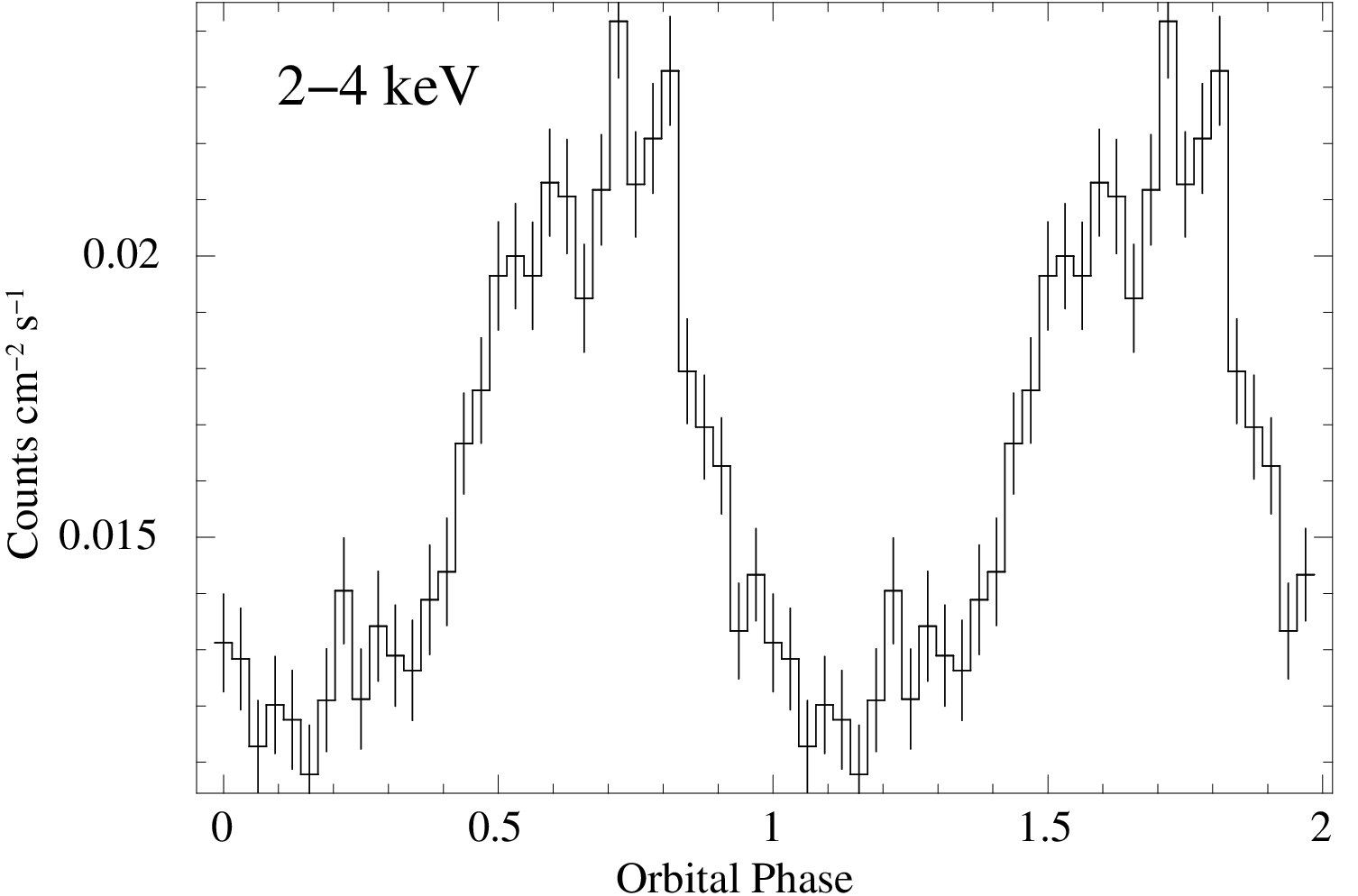} 
\includegraphics[angle=-90,scale=0.25]{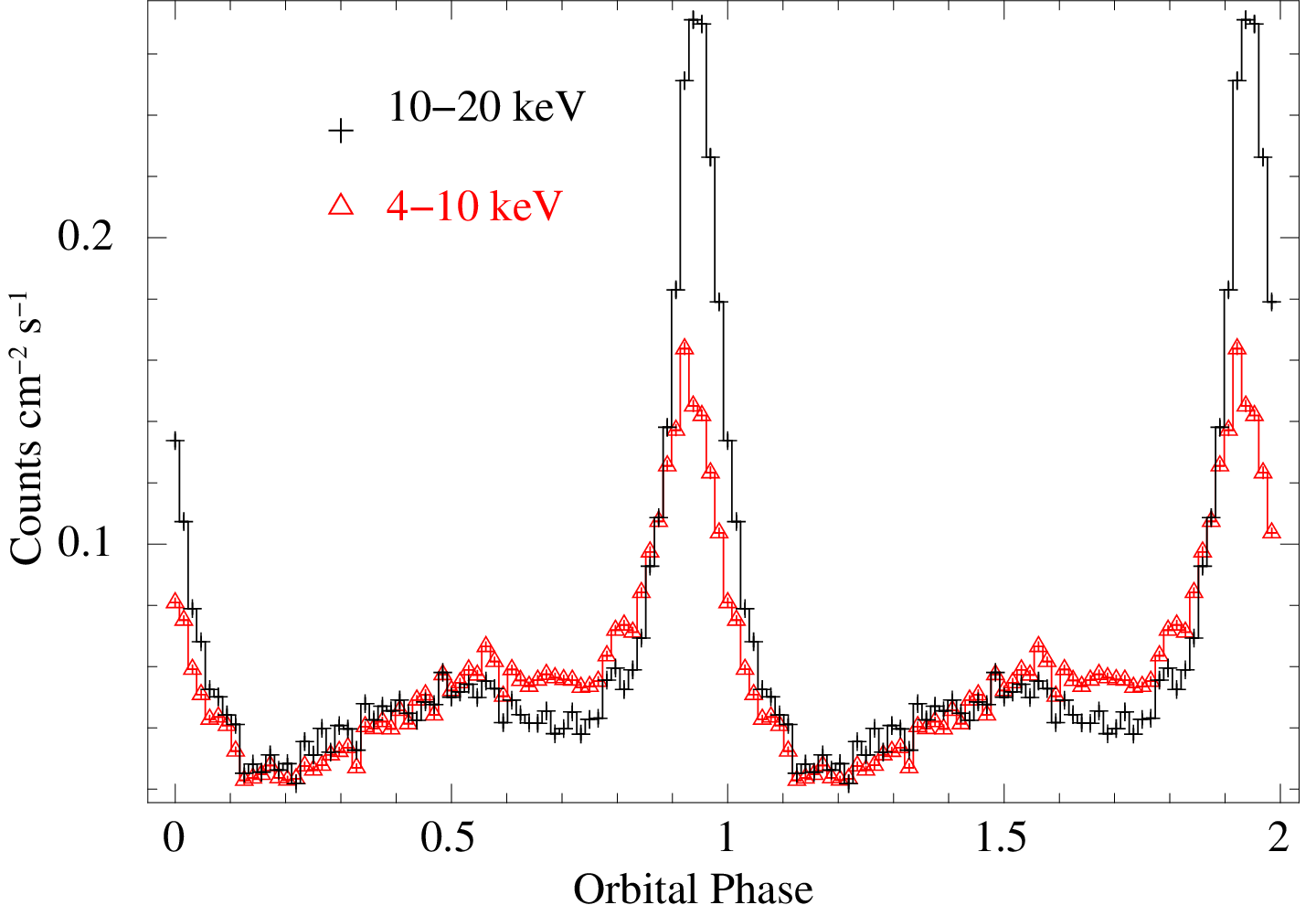} \\[1ex]
\includegraphics[angle=-90,scale=0.25]{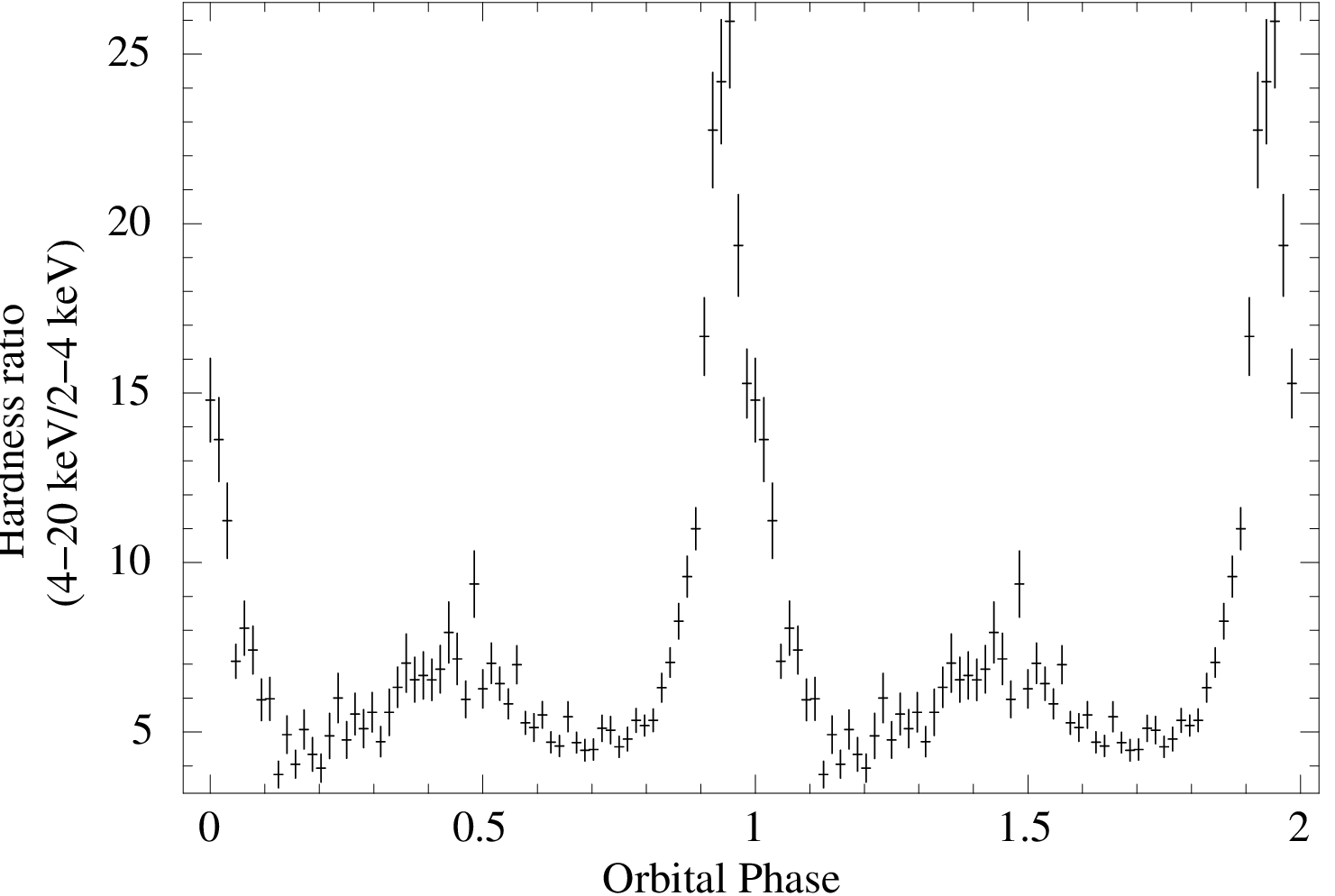} 
\includegraphics[angle=-90,scale=0.25]{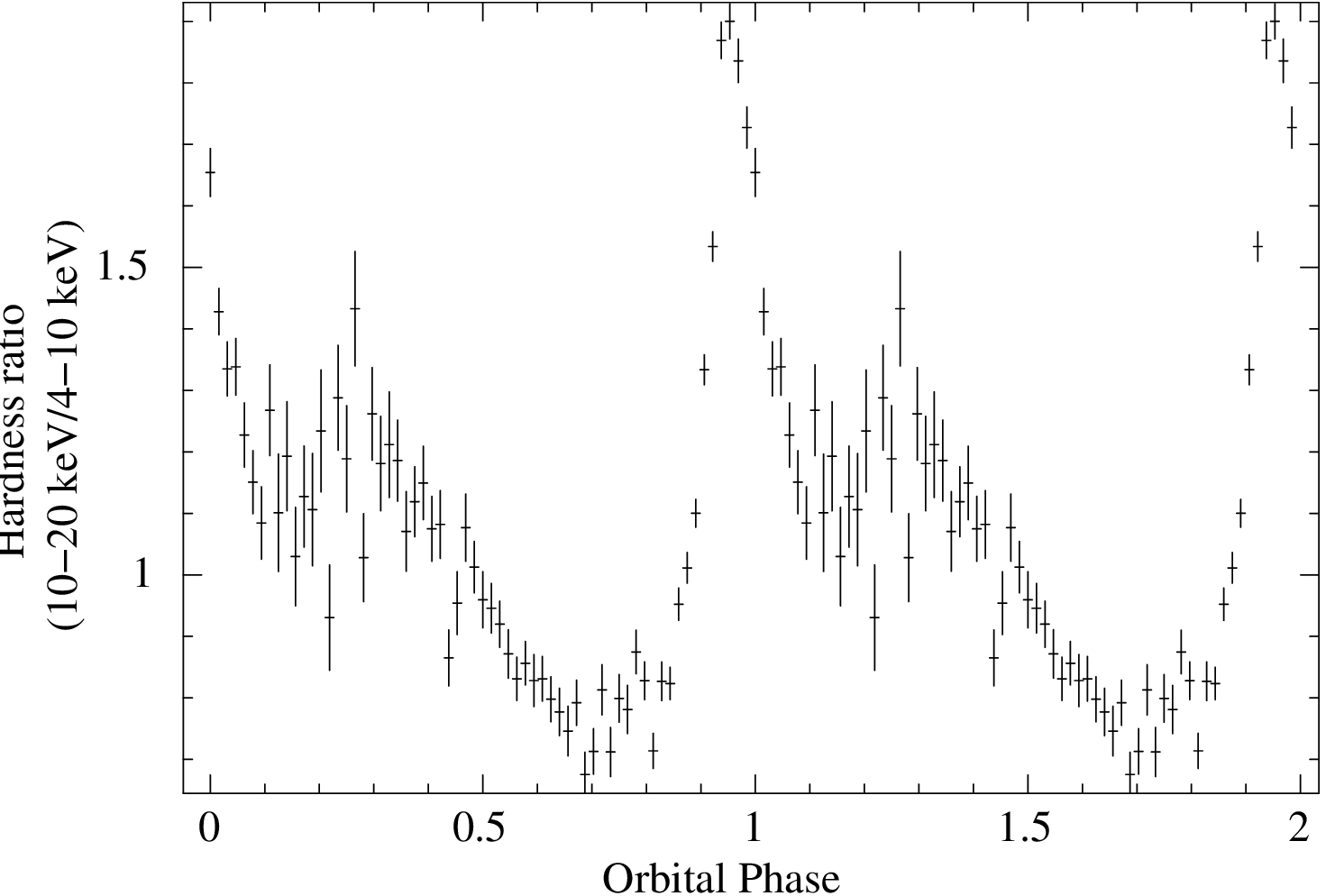}
\caption{The top left panel is the orbital intensity profile of GX 301-2 in 2-4 keV energy band of {\it MAXI}. The top right panel is the orbital intensity profile in 
4-10 keV (triangles) and 10-20 keV (cross) energy band. The bottom left panel is the hardness ratio (HR) of the 
counts in 4-20 keV energy band to low energy 2-4 keV band. Lower right panel is the hardness ratio of counts in 10-20 keV and 4-10 keV energy band.}
\label{pulseprofiles}
\end{figure*}

We have first used 16 years long light-curve of GX 301-2 obtained with the {\it RXTE ASM} in 2-12 keV energy band \citep{levine1996} to estimate the orbital period 
of the system (P$_{orb}$=3584304 s; consistent with \cite{koh1997}). We have then folded the {\it MAXI-GSC} light curves of GX 301-2 obtained during
MJD:55064 to 56502  with the orbital period obtained from {\it RXTE ASM} to obtain energy resolved orbital intensity profiles. The orbital phase reference is taken 
from \citet{koh1997}, with the phase zero corresponding to periastron passage. The orbital intensity profiles in the {\it MAXI} energy bands of 2-4 keV, 4-10 keV and 
10-20 keV are shown in Figure.~\ref{pulseprofiles} along with hardness ratios (4-20 keV / 2-4 keV and 10-20 keV / 4-10 keV) of photons. The orbital modulation 
is most pronounced in the highest energy band of 10-20 keV showing the pre-periastron X-ray flare as well as the intermediate intensity 
apastron peak. In the lowest 2-4 keV energy band, instead of a strong flare, a sinusoidal modulation in intensity can be seen. This smoothly varying soft X-ray
profile is seen for the first time in GX 301-2 with the {\it MAXI} data. As seen in Figure.~\ref{pulseprofiles}, there is an increase in the 4-10 keV photons after 
the apastron passage, which is reflected in the low hardness ratio values around orbital phase 0.5-0.8.

\subsection{Orbital phase averaged and phase resolved spectroscopy}

\subsubsection{Orbital phase averaged spectrum}

\begin{figure*}
\centering
\includegraphics[angle=-90,scale=0.3]{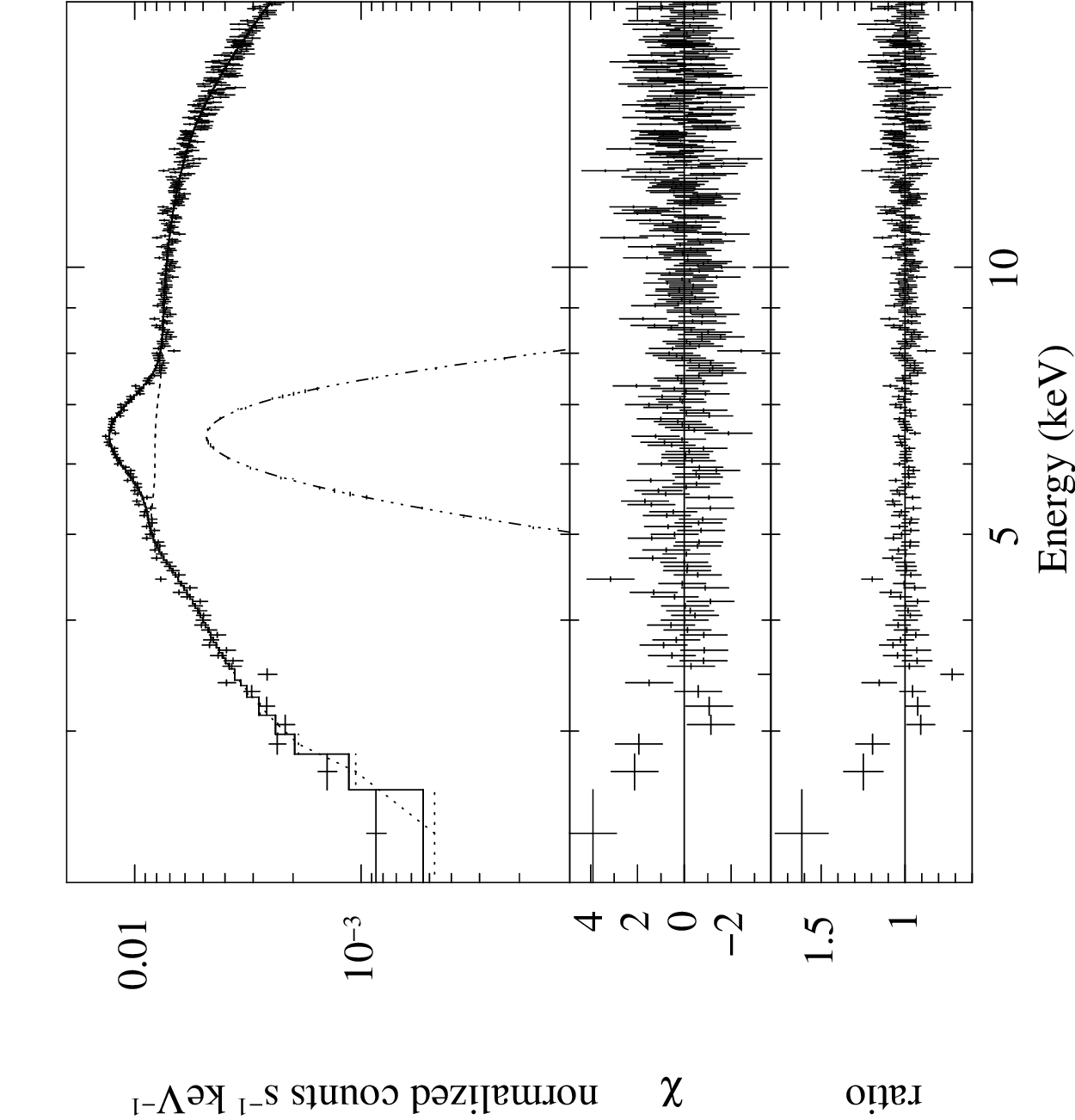}
\hspace{1.5cm}
\includegraphics[angle=-90,scale=0.3]{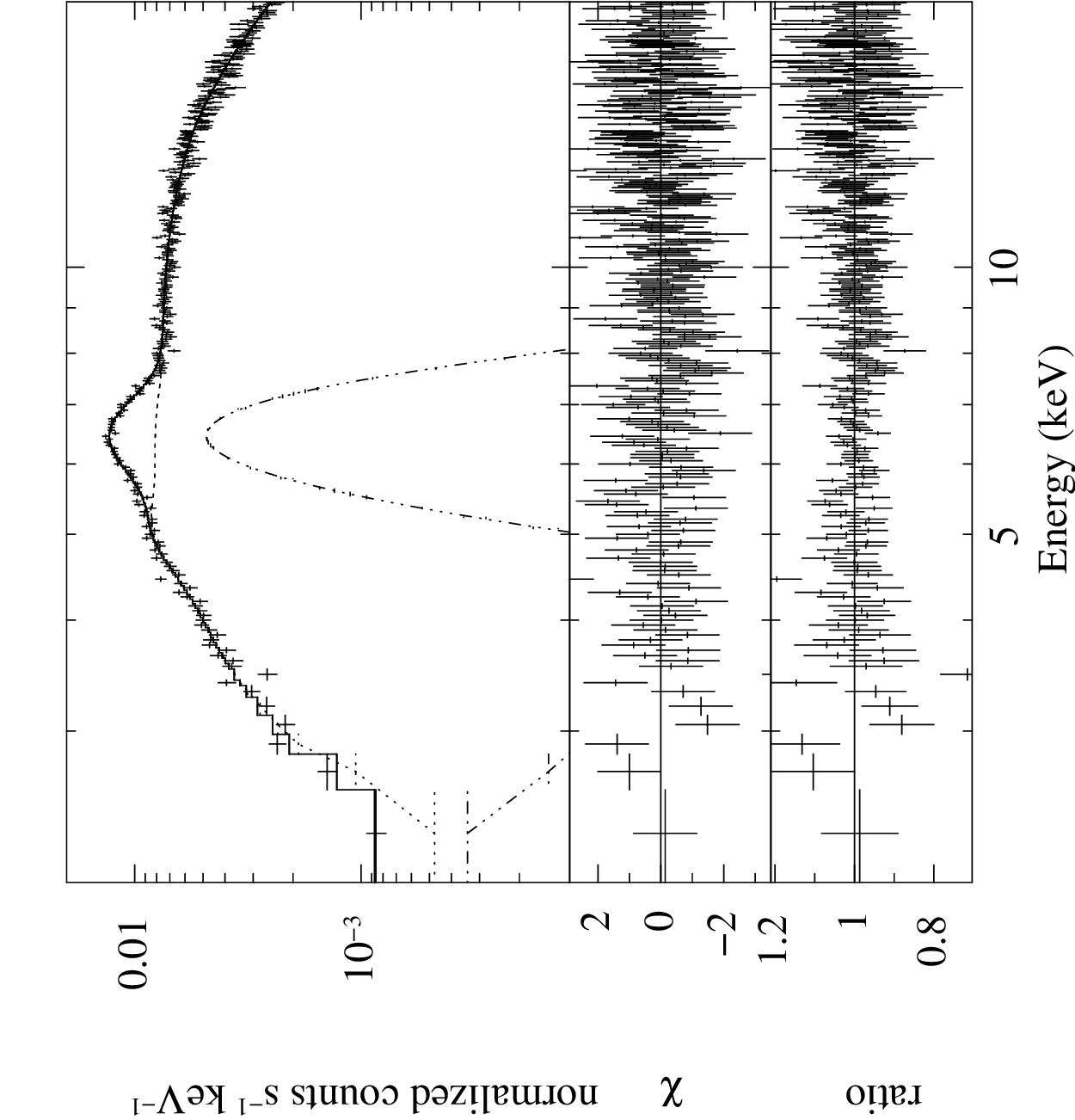}
\caption{The orbital phase averaged X-ray spectrum is shown along with the best fit model components (top panel), contributions of the residuals to the 
chisquare (middle panel) and ratio between data and the model (bottom panel) in the 2-20 keV band. The best fit model is obtained in the energy range 3.5-20 keV 
and is extended to 2 keV. The left panel shows the presence of a soft excess and the right panel shows the fit after modelling the soft excess with a blackbody 
component.}
\label{phase_avg}
\end{figure*}

We have extracted the orbital phase averaged spectrum of GX 301-2 with {\it MAXI GSC} for the same observation duration using the {\it MAXI} on-demand data 
processing \footnote{http://maxi.riken.jp/mxondem/}. The orbital phase averaged spectrum is modelled with a power-law with a high energy cut-off 
modified by photo-electric absorption by column density of absorbing material along our line of sight. The spectral analysis is performed using {\it Xspec v:12.6}. 
Though the broadband X-ray spectrum of GX 301-2 is known to have a partial covering absorption \citep{mukherjee2004,suchy2012,labarbera2005}, in the 
limited energy band of the GSC, it was not required to include partial covering absorption in the spectral model. A Fe K$\alpha$ fluorescence line present in 
the spectrum is modelled with a single gaussian line. The analysis is carried out by fitting the spectrum from energy range 3.5 keV to 20 keV by a power-law 
with a high energy cut-off and a single gaussian line. When this fit was extended to lower energies till 2.0 keV, a low energy excess was found in the spectrum. 
This low energy excess was modelled by an unabsorbed blackbody component by freezing all other parameters of the previous best fit model. The motive behind 
modelling the low energy excess by an unabsorbed blackbody is to only estimate the flux of the low energy excess and not to derive a blackbody temperature 
and size of the region. The orbital phase averaged X-ray spectrum along with the best fit model components is shown in Figure.~\ref{phase_avg}, along 
with the presence of soft excess, which is then modelled by a blackbody component. 
The $\chi^{2}_{\nu} $ is 0.99 for 352 degrees of freedom.

\subsubsection{Orbital phase resolved spectra}

Strong variations of the hardness ratio of photons in 10-20 keV energy band and 4-10 keV energy band with the orbital phase shown in
Figure.~\ref{pulseprofiles} indicates significant changes in the overall spectrum. Many observations of GX 301-2 have been carried out at different orbital 
phases in the past \citep{endo2002,mukherjee2004,labarbera2005,suchy2012}, but with uneven phase coverage. We have performed an orbital phase resolved 
spectral analysis of GX 301-2, by using the arbitrary grouping of the individual scans allowed by {\it MAXI} on demand data processing \citep{nakahira2012,doroshenkov2013}.
Using the orbital ephemeris, we extracted the start time and end time for different orbital phase bins, and extracted the spectra for each orbital
phase with data from multiple orbital cycles by grouping the individual scans. We have chosen 21 independant orbital bins, with the bin size
chosen such that it can constrain the Fe line parameters. The effective exposure times varies from 43 kilosec for the orbital bins near the X-ray peak to 454 kilosec
for orbital bins near dim phase. \\
In this work, we have fitted these 21 orbital phase resolved spectra with two models: a power-law continuum, with and without a high energy 
cut-off. For the spectral fits with a high energy cut-off, the value of cut-off energy and fold energy (E$_{c}$ and E$_{f}$) was frozen to 
the orbital phase averaged values of 15 keV and 18.6 keV respectively, since the parameters of the high energy cut-off cannot be constrained well in the phase 
resolved spectra. A Fe fluorescence line was found in all the orbital phases which was modelled by a single gaussian line. For both the spectral models, 
the spectra were fitted from energy range 3.5 keV to 20 keV and then the fit was extended to lower energies till 2.0 keV. 
For some orbital phases near the X-ray peak, a low energy excess is found to be present in the spectra. To only estimate the flux in the soft excess, 
we have modelled the low energy excess with an unabsorbed blackbody component. The range of $\chi^{2}_{\nu}$ values, after accounting for the low energy excess), 
is 0.8-1.16 (0.75-1.23) for a power-law with high energy cut-off (without the high energy cut-off). \\

\begin{figure*}
\centering
\includegraphics[angle=-90,scale=0.4]{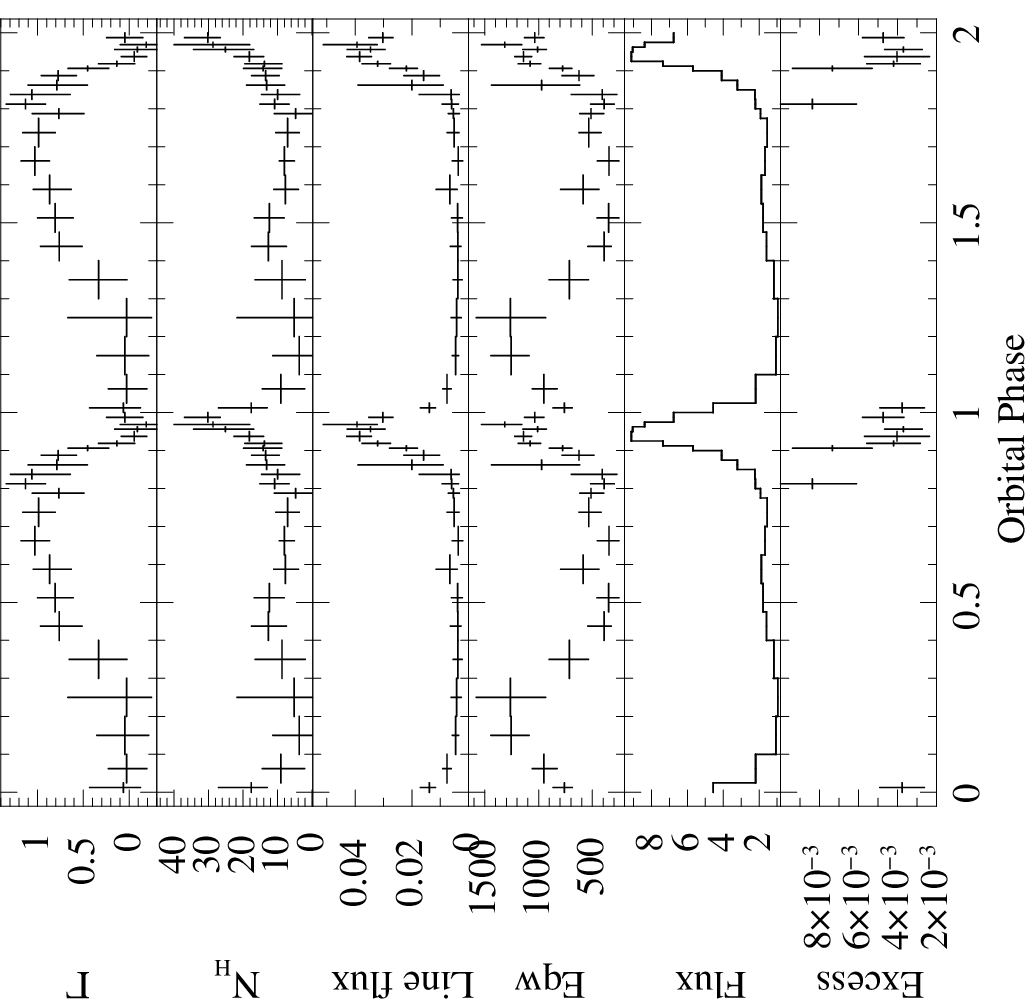}
\hspace{1.2cm}
\includegraphics[angle=-90,scale=0.4]{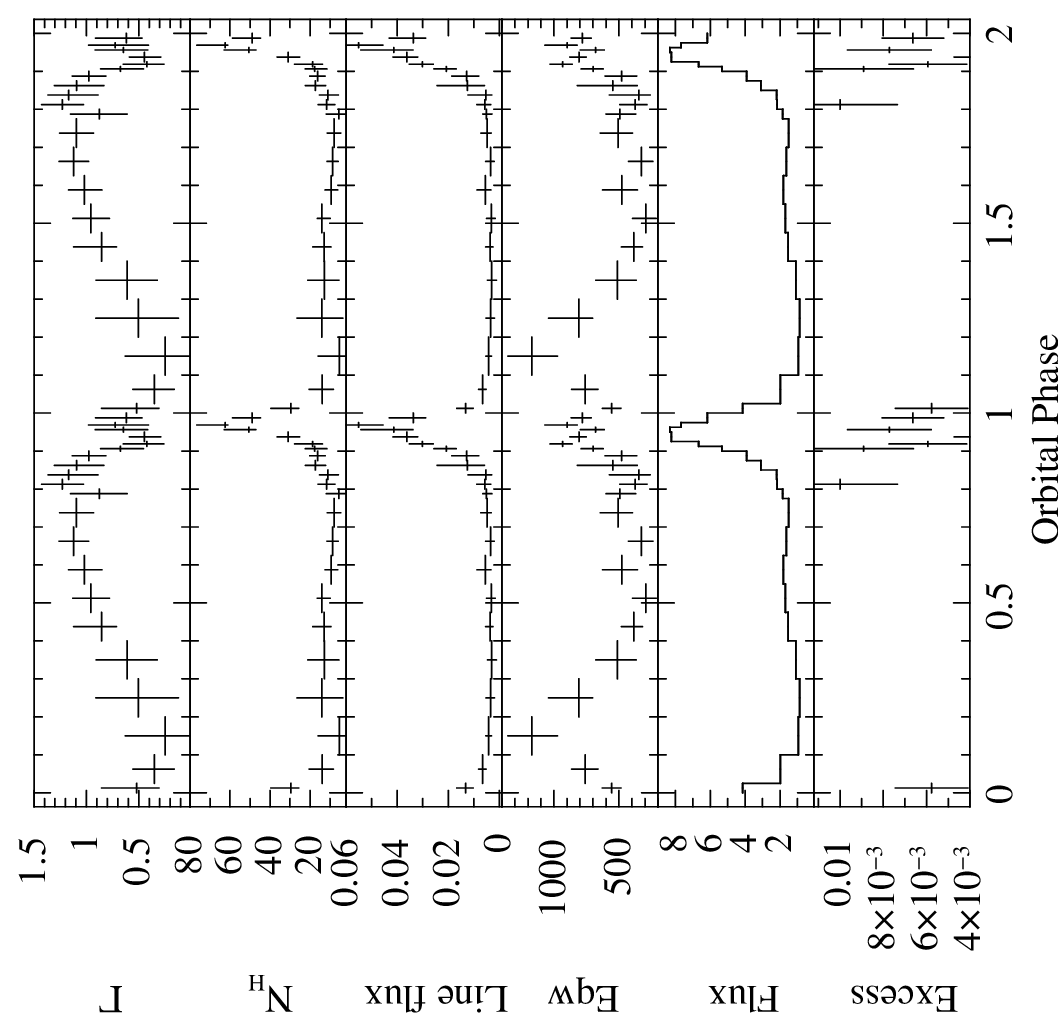}
\caption{Variation of Photon index ($\Gamma$), equivalent column density of hydrogen (N$_{H}$ in units of 
$10^{22}$ cm$^{-2}$), Line flux of Fe K$\alpha$ (in units of photons cm$^{-2}$ s$^{-1}$), Equivalent width of Fe K$\alpha$ line (Eqw in units of eV),
Flux of source (F in the units of $10^{-9}$ ergs s$^{-1}$ cm$^{-2}$) and the ratio of flux in the low energy excess modelled by a blackbody to the 
flux of the system(Excess) for both the spectral models: power-law with a high energy cut-off (left panel) and a power-law (right panel).}
\label{parameters}
\end{figure*}

The orbital dependance of the spectral parameters photon index $\Gamma$, column density of absorbing matter along our line of sight N$_{H}$, flux and equivalent 
width of Fe line emission, total flux of the system in the entire 2-20 keV energy band of {\it MAXI GSC} and ratio of flux included in the low excess to the total flux 
of the system, for both the spectral models are shown in the Figure.~\ref{parameters}. The errors on the spectral parameters $\Gamma$, N$_{H}$ and flux of Fe line 
emission are at 90\% confidence level whereas the errors on equivalent width of Fe line is shown at 1$\sigma$ confidence level. The errors on the relative flux 
in the low energy excess, modelled by a blackbody is taken at 1$\sigma$ confidence level of blackbody normalisation. Since the high energy cut-off affects the 
X-ray spectrum only above 15 keV, therefore the two results shown in Figure.~\ref{parameters} with and without the high energy cut-off included in the spectral 
model show similar pattern of variation of the spectral parameters. \\
The orbital variation of column density of absorbing matter along our line of sight N$_{H}$ and flux of Fe line emission for both the spectral models follows the 
orbital variation of the flux of the system. However, orbital variation of power-law photon index and equivalent width of Fe line follows a different pattern from 
the flux of the system. The power-law photon index $\Gamma$ is hard at the dim phase and at the pre-periastron X-ray flare and is soft around the apastron passage. 
The orbital variation of equivalent width has two peaks and the lowest equivalent width occurs at the apastron passage. The soft excess is detected only near the 
X-ray flare and has a flux that is less than 1.0\% of the continuum flux.\\

\begin{figure*}
\centering
\includegraphics[angle=-90,scale=0.3]{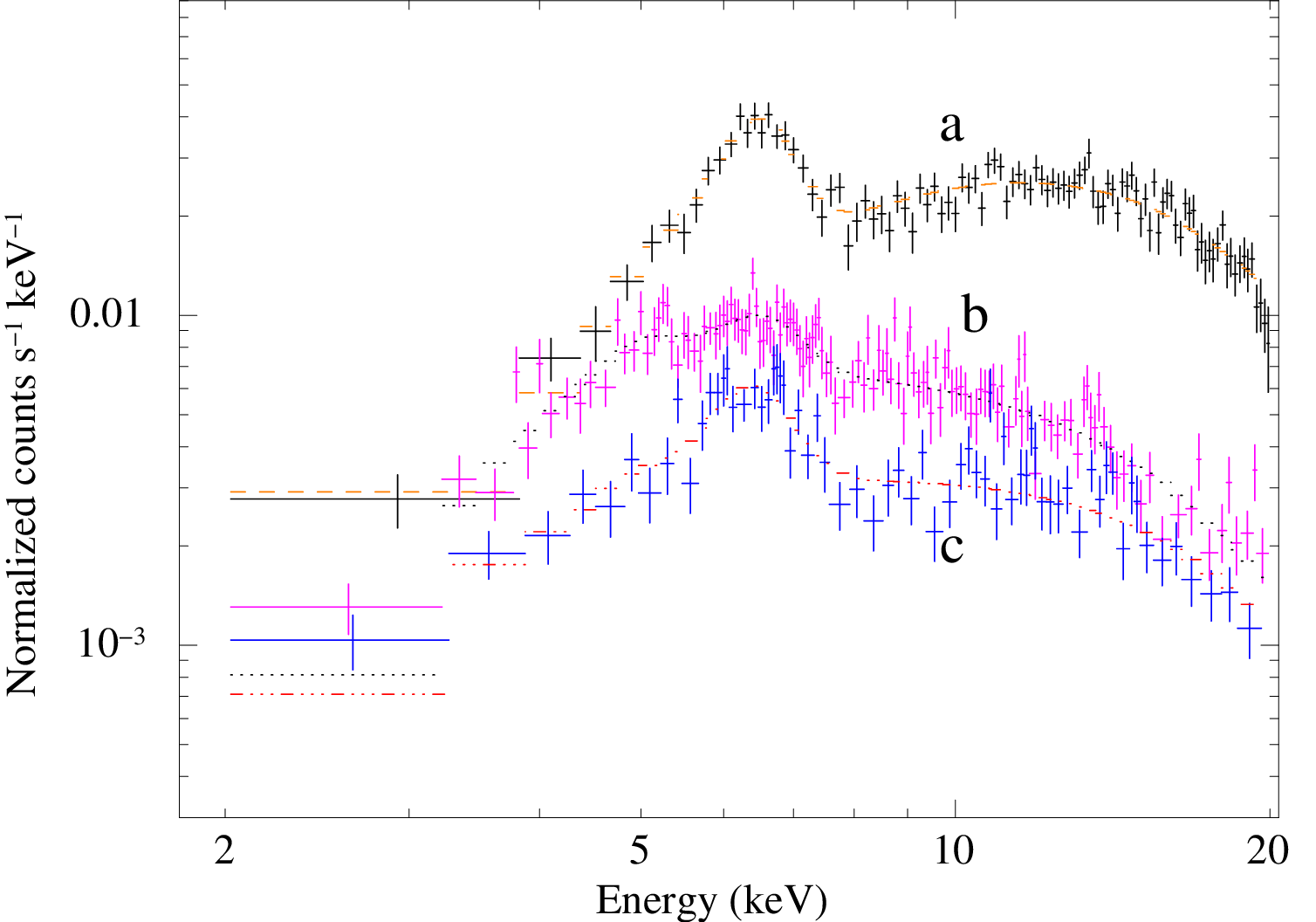}
\includegraphics[angle=-90,scale=0.3]{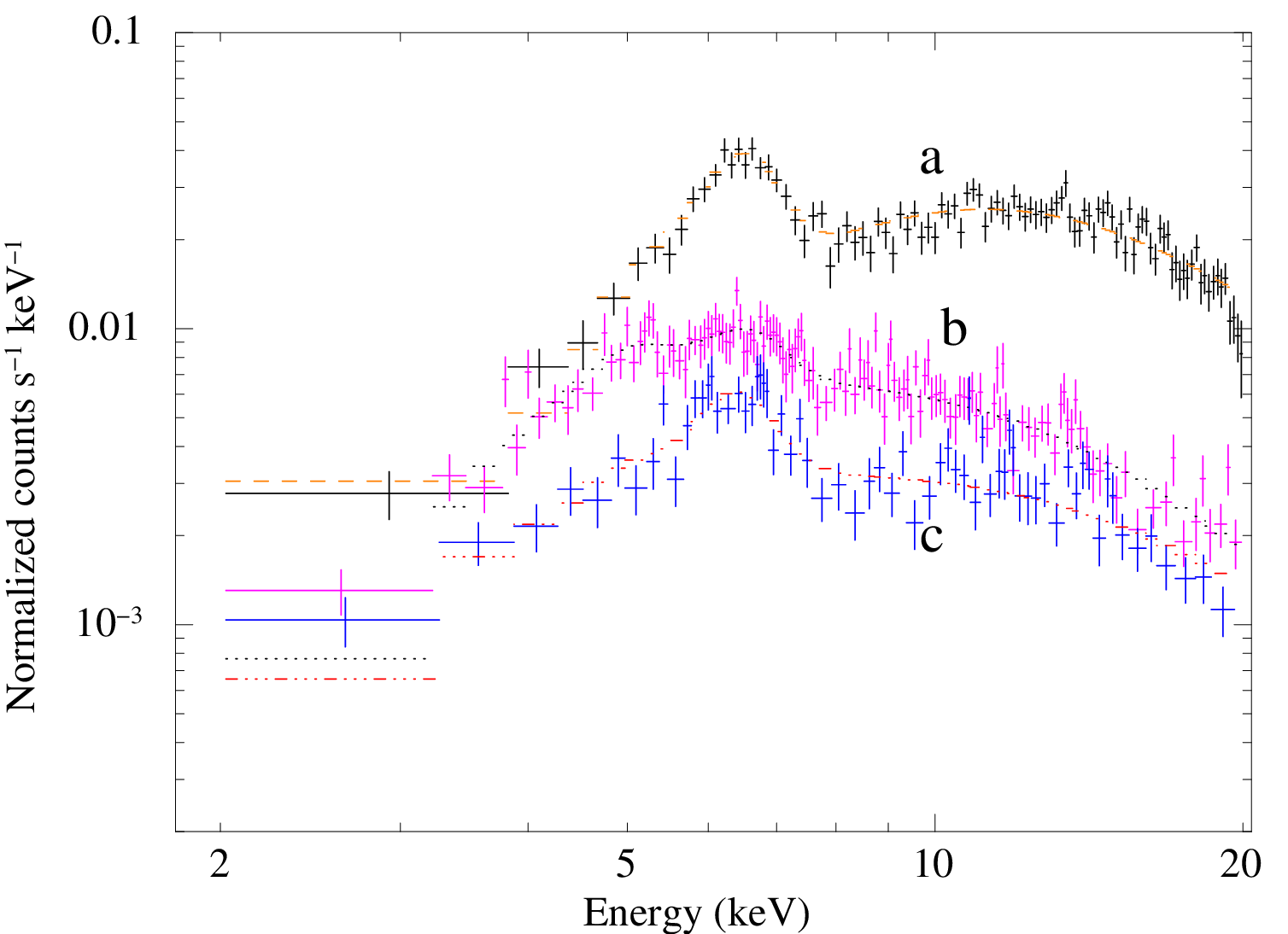}
\caption{The spectra and the best fitted model are shown here for three different orbital phases: the pre-periastron X-ray flare (a), 
the apastron passage which has the lowest equivalent width of the iron line (b) and the dim phase which has the highest width of the iron line (c), 
for both the spectral models, i.e power-law with a high energy cut-off (left panel) and power-law model(right panel).}
\label{rep_setpadd}
\end{figure*}

Three of the phase resolved X-ray spectra showing the highest equivalent width of Fe line (dim phase : 0.1-0.2), lowest equivalent width (intermediate phase: 
0.475-0.55) and at pre-periastron flare (phase 0.95) are shown in Figure.~\ref{rep_setpadd}.

\section{Discussions}

\subsection{Multi-band orbital modulation}

Figure.~\ref{pulseprofiles} shows the orbital intensity profiles of GX 301-2 in 2-4 keV, 4-10 keV and 10-20 keV energy bands of {\it MAXI}, along with the hardness 
ratios of photons in 4-20 keV energy band to 2-4 keV energy band and 10-20 keV energy band to 4-10 keV energy band. 
We have detected a strong, nearly sinusoidal, modulation by a factor of $\sim$ 2.5 in the 2-4 keV energy band, the profile of which is different from the 
orbital modulation in the 4-20 keV band. In the entire 4-20 keV band and other sub-bands (4-10 keV and 10-20 keV), the source shows an appreciable orbital 
modulation, as previously known, with a strong pre-periastron X-ray flare as well as an apastron peak. The hardness ratio between the 10-20 and 4-10 keV bands 
shows the pre-periastron X-ray flare to be hard which is also reflected in small value of the power-law photon index at the phase $\sim$ 0.95 
(Figure.~\ref{parameters}). After the apastron passage and before the hard X-ray flare (orbital phase range 0.55-0.85), there is a decrease seen in the hardness 
ratio. A corresponding high value of the power-law photon index $\Gamma$ is seen in the Figure.~\ref{parameters} .

\subsection{Variation of Spectral parameters with Orbital Phase}

We have studied the orbital dependence of different spectral parameters of GX 301-2 using a power-law continuum model with and without a high energy cut-off. 
Figure.~\ref{parameters} shows the orbital variation of $\Gamma$, N$_{H}$, flux and equivalent width of Fe fluoresence line, total flux of the system and 
ratio of flux included in the low excess to the total flux, for both the spectral models. From Figure.~\ref{parameters}, the orbital variation in spectral 
parameters for both the models show very similar trend. The photon index gradually increases from a low value of 0.05 for power-law with high energy cut-off 
(0.3 for power-law) model at the dim orbital phase 0.1-0.3, to 1.1 for both the models at the periastron passage, just before the X-ray flare. 
The spectrum is found to be hard at dim phase 0.1-0.3 for both the models. Spectral hardening at this phase was reported in earlier works using data from 
{\it RXTE-ASM} and {\it SuperAGILE} \citep{pravdo2001,evangelista2010}. \\
The column density N$_{H}$ is found to vary with a pattern similar to the flux of the system, indicating a possible origin of flare due to increased mass 
accretion at the orbital phase of $\sim$ 0.95. The absorption column density is in the range of 10-20 $\times 10^{22}$ cm$^{-2}$ during most of the orbit, except 
near the flare. 
Since the data from {\it MAXI} is averaged over multiple orbital cycles, clumpiness in wind as seen in {\it RXTE PCA} data over one cycle \citep{mukherjee2004} 
and {\it BeppoSAX} data over one cycle \citep{labarbera2005}, is not detectable in the {\it MAXI} data and a partial covering absorption is not required to 
fit the X-ray spectra obtained with {\it MAXI GSC}. The flux of Fe line emission also shows a strong increase during the flare, by a factor of $\sim$ 3, compared 
to the nearly constant flux in the orbital phase range 0.1-0.8. \\
The orbital variation of equivalent width of Fe line shows a markedly different trend as compared to the orbital variation of column density and total flux of the system. 
The highest equivalent width occurs at the dim phase of 0.1-0.3 which also has lowest N$_{H}$ along the line of sight. This different pattern of 
orbital variation of equivalent width and the column density N$_{H}$ provides us important clues about the distribution of matter around the system, 
which is discussed further in the next section. \\
A low energy excess, contributing to less than 1.0\% of the continuum flux, is found in the spectra at orbital phases near 
the pre-periastron X-ray flare as shown in the bottom panels of Figure.~\ref{parameters}. 
Low energy excesses in the X-ray spectrum of GX 301-2 has been previously reported by \cite{haberl1991} using {\it EXOSAT} observation and \cite{saraswat1996} using 
{\it ASCA} observation. Due to high value of N$_{H}$ around the X-ray flare, we rule out the possible origin of low energy excess from the surface of the neutron star 
\citep{hickox2004}. A possible origin of this low energy excess could be due to the X-ray shocks in the system when the high density accretion stream interacts 
with the stellar wind from WRAY 977 \citep{haberl1991,kaper2006}.

\subsection{Orbital modulation: wind diagnostics}
\subsubsection{Column density as tracer}

\begin{figure*}
\centering
\includegraphics[angle=-90,scale=0.3]{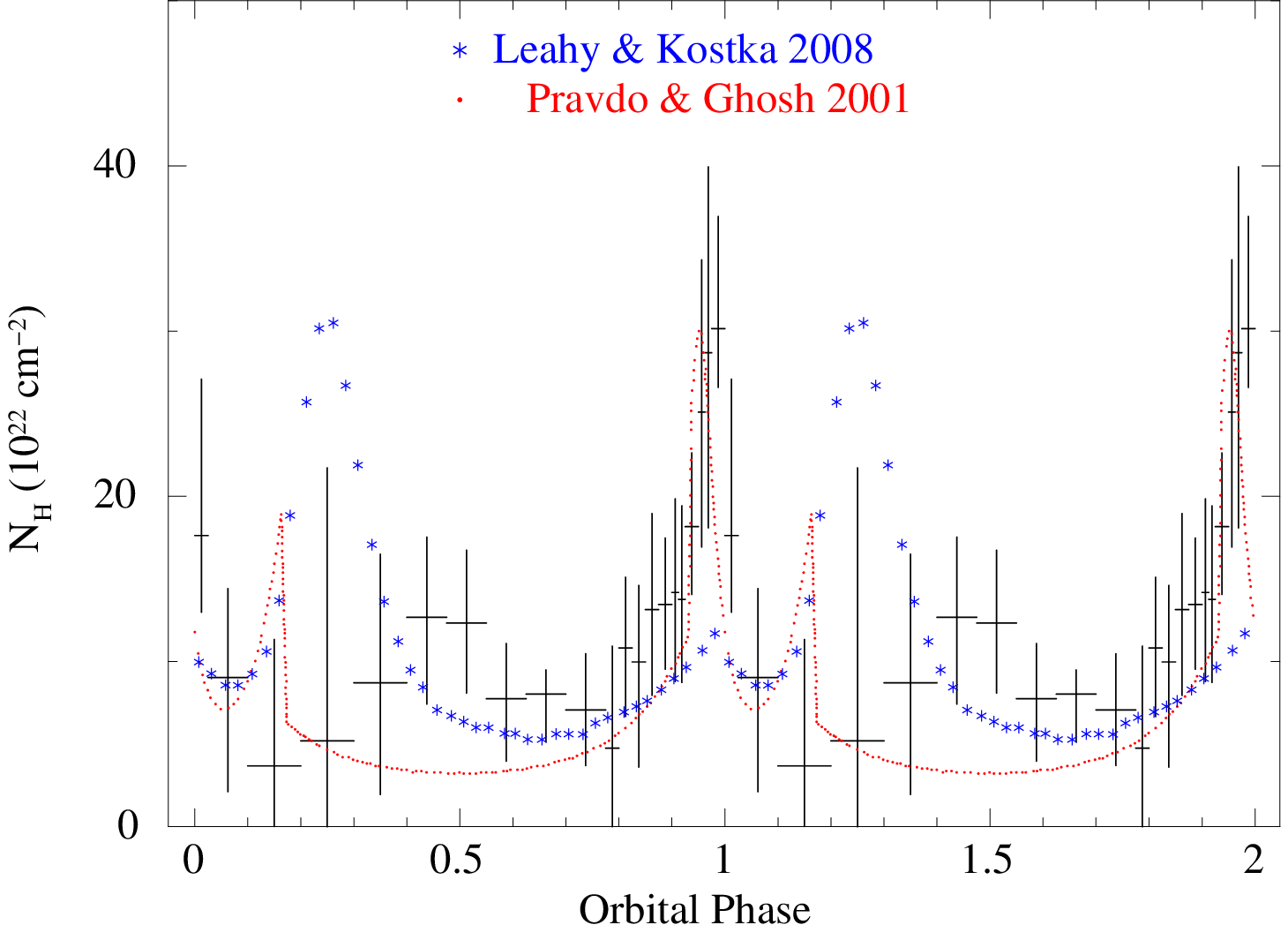}
\includegraphics[angle=-90,scale=0.3]{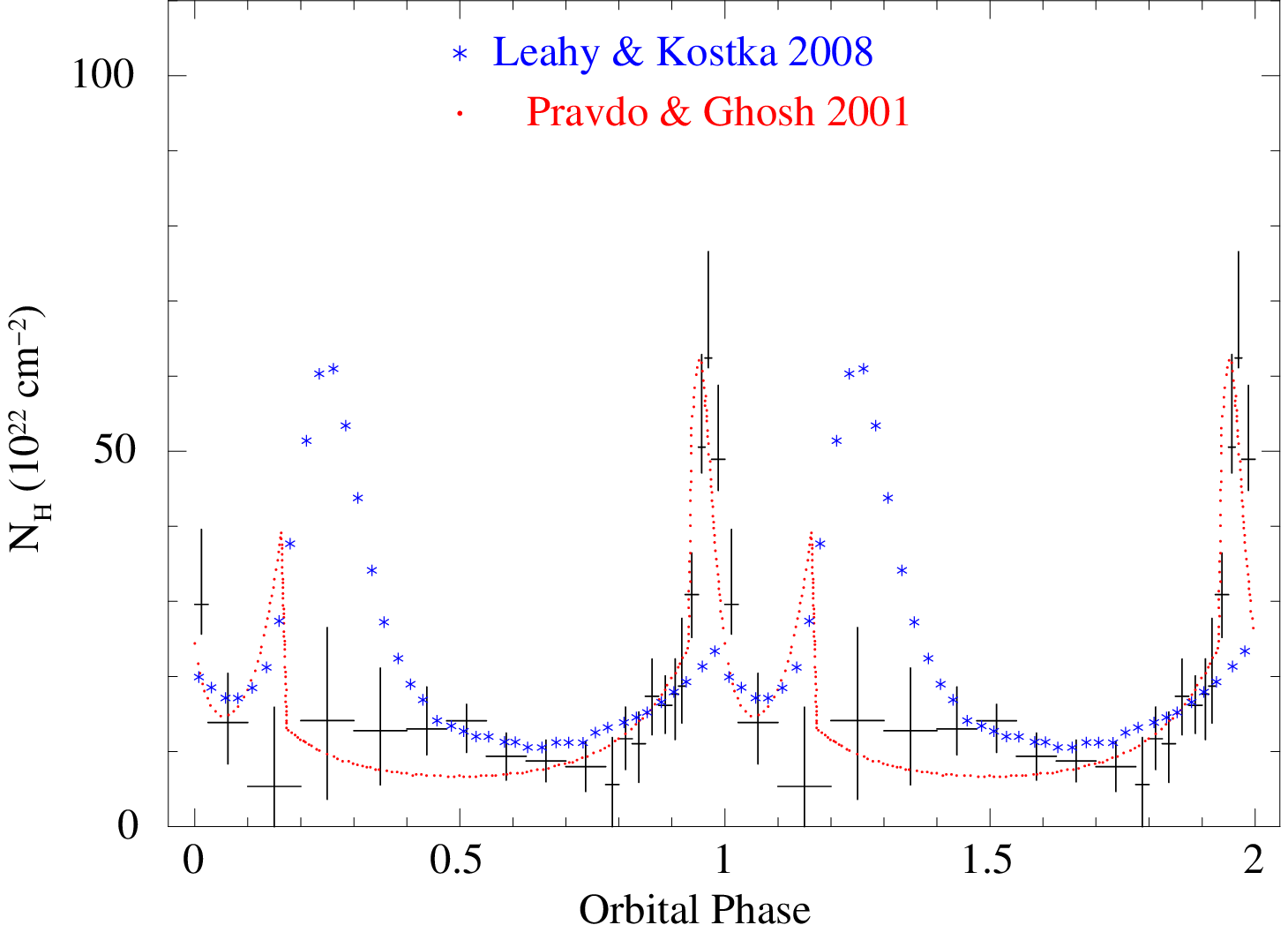}
\caption{Variation of column density N$_{H}$ as function of orbital phase of GX 301-2 for both power-law with high energy cut-off (left panel) and power-law model 
(right panel), overlaid on the predicted orbital variation of column density (circles - Pravdo \& Ghosh (2001), asterisks - Leahy \& Kostka (2008)). The 
predicted orbital variation of column density is normalised to the highest column density found in our results for the sake of comparing with the observed 
orbital profile.}
\label{nH_compare}
\end{figure*}

Figure.~\ref{nH_compare} shows the orbital variation of the column density of absorbing matter along our line of sight N$_{H}$ obtained in this work with 
{\it MAXI} for both a power-law with a high energy cut-off model and a power-law model and is compared to the predicted orbital variation of N$_{H}$ expected 
from \cite{pravdo2001} and \cite{leahy2008} models. These models were developed to describe the orbital intensity profile of GX 301-2 as seen with 
{\it BATSE} \citep{pravdo2001} and {\it RXTE ASM} \citep{leahy2008}. For the sake of comparison with the measured orbital profile of the column density, we have 
normalised the highest column density predicted by these models to the highest observed column density. \\
The \cite{pravdo2001} model predicts an enhanced column density twice, at the pre-periastron X-ray flare and at the dim orbital phase. However, only one 
peak in column density is observed with {\it MAXI} during pre-periastron passage. The \cite{leahy2008} model underestimates the value of column density at the
pre-periastron phase but predicts a large increase in column density around phase 0.2. On the other hand, the {\it MAXI GSC} spectra do not show an enhanced 
column density at orbital phase 0.1-0.3 predicted by both the models, in our results. A possible explanation for this would be that the absorbing matter 
considered in these two models is situated away from the line of sight at this phase range. To further investigate this possibility, we have used the orbital 
variation of the equivalent width of the fluorescent iron line in the next section as an additional observational signature to study the orbital distribution of 
circumstellar matter around the neutron star in GX 301-2.

\subsubsection{Iron line as tracer}

The iron fluorescence line provides important information on the geometry and density of the line producing region around the central X-ray source that produces 
the continuum X-ray emission \citep{makishima1986}. In HMXBs, the fluorescence lines originate from the dense absorbing material present in the stellar wind. 
The detection of a Compton shoulder of Fe K$\alpha$ in GX 301-2 by \citet{watanabe2003} confirms the presence of Compton thick matter around the X-ray source. 
The equivalent width of the Fe line is expected to be linearly correlated to the column density for an isotropically distributed cold matter 
\citep{inoue1985,kallman2004}. Deviation from this linear correlation can be mostly attributed to any anisotropic configuration of the 
reprocessing matter around the X-ray source. \\
\begin{figure*}
\centering
\includegraphics[angle=-90,scale=0.3]{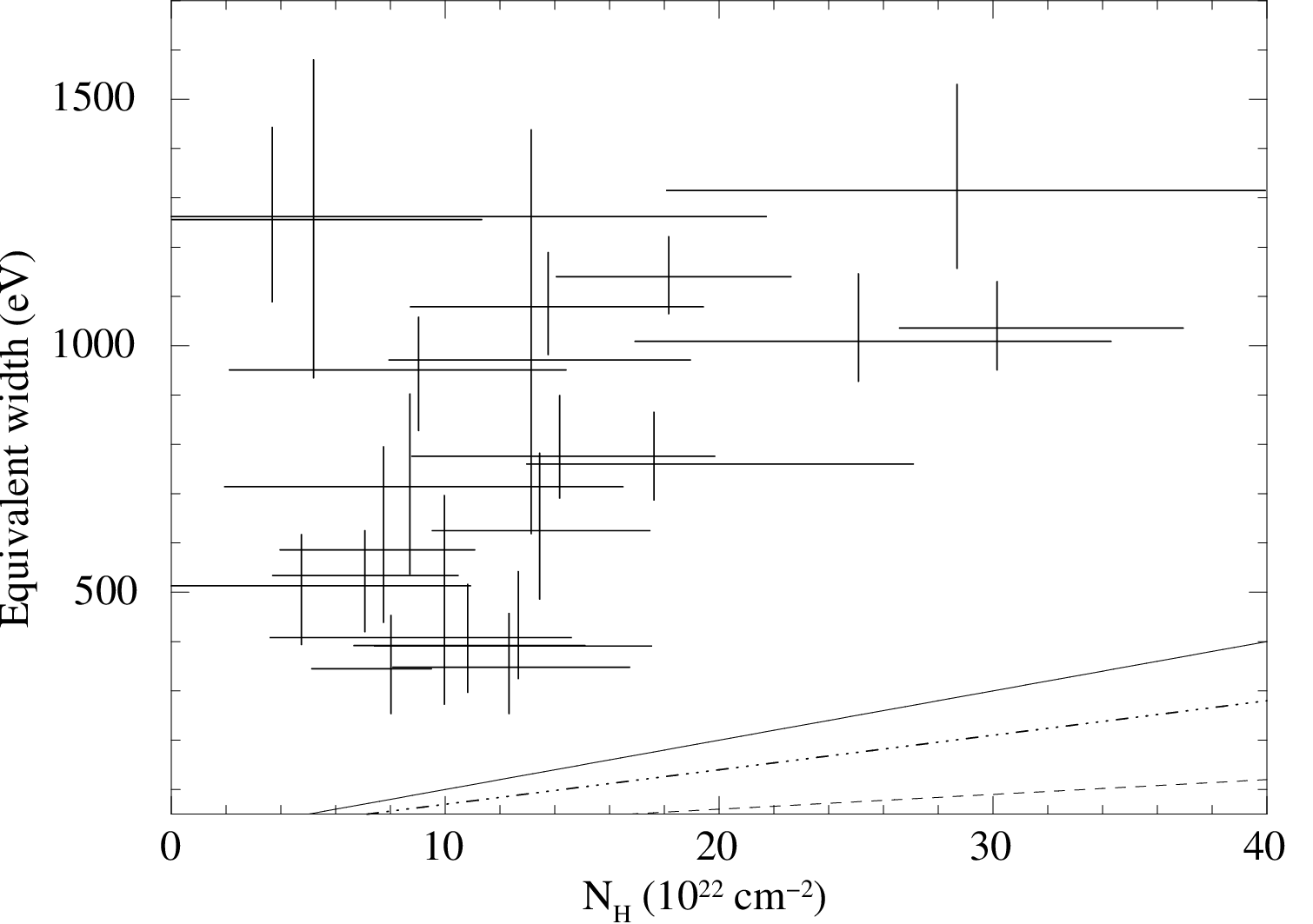}
\includegraphics[angle=-90,scale=0.3]{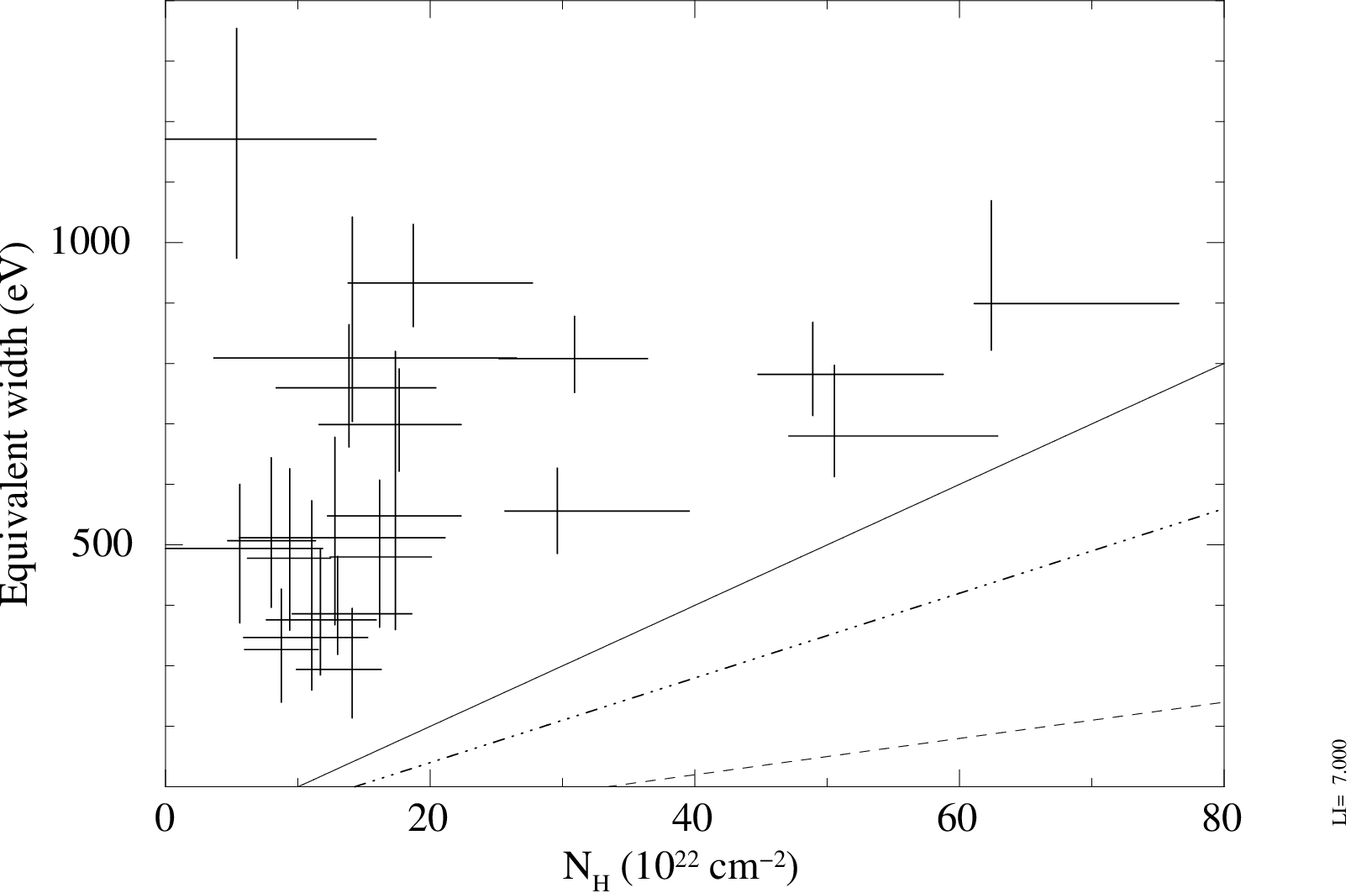}
\caption{Plot of equivalent width of Fe K$\alpha$ versus absorbing column density N$_{H}$ obtained for different orbital phases with MAXI 
for both power-law with highecut (left panel) and power-law model (right panel). Solid line represents the 
relation between equivalent width and column density of absorbing matter for isotropically distributed matter for $\Gamma \sim$ 1.1 calculated by 
Inoue (1985). The dashed line and dot dashed line is the relation between equivalent width and column density for a spherical shell of gas for 
$\Gamma \sim$ 1 and $\Gamma \sim$ 0.2 respectively calculated by Kallman et al. (2004). $\Gamma \sim$ 0.2 corresponds to the hardest spectrum seen in the dim phase 
with MAXI, where the highest value of equivalent width is found.}
\label{nH_eqw}
\end{figure*}

Figure.~\ref{nH_eqw} shows the plot of equivalent width of the iron line against the absorbing column density N$_{H}$ in different orbital bins for GX 301-2 in 
this work for both the spectral models. In certain orbital phases, we do see very high equivalent width ($\sim$ 1 keV) along with a small value of column 
density ($\sim$ 2-5 $\times 10^{22}$ cm$^{-2}$). These observations highly deviate from the relation expected for an isotropically distributed gas 
\citep{inoue1985,kallman2004} and instead there seems to exist high anisotropicity in the distribution of circumstellar matter around the X-ray pulsar, 
especially in some orbital phases. Our results are in contrast to the studies by \citet{makino1985,endo2002,furst2011}, where they found a linear correlation 
between equivalent width and column density, thereby favouring a spherical shell model of gas distribution. However, such studies were carried out in narrow 
window of pointed observations and this work is carried by averaging over multiple orbital cycles. \\
From orbital variation of N$_{H}$ and equivalent width shown in Figure.~\eqref{parameters}, we see a very high equivalent width for very low line of sight 
column density at orbital phases 0.1-0.3. 
This being an important outcome of the present work and the fact that it is hardest to measure absorption column density in the dim phase, we have further 
examined it. As the MAXI-GSC spectrum has very little count-rate below 3 keV, we carried out the spectral analysis above 3.5 keV and an upper limit on N$_{H}$ of 
11 $\times 10^{22}$ cm$^{-2}$ is obtained at the orbital phase 0.1-0.3 with a 90\% confidence limit. Near the flare phase, the spectrum has higher statistics 
and the best fit model for the spectrum above 3.5 keV, when extended to the lower energy band, clearly shows the presence of a soft excess. The blackbody 
temperature of the soft excess measured in these phases is in the range of 0.1-0.3 keV and this component does not have any significant contribution in the 
GSC spectrum above 3.5 keV. Therefore, the value of N$_{H}$ measured at different orbital phases (especially at dim phase of 0.1-0.3) represents the true line 
of sight value of N$_{H}$. A possible explanation would be that the distribution of circumstellar matter in these orbital phases somehow 
avoids the line of sight and the possible geometries could be the ones in which the reprocessing region (where the Fe fluorescence line is produced) is 
situated behind the neutron star or on one side of the neutron star away from the line of sight. The optical studies of GX 301-2 done by \cite{kaper2006} 
confirms the presence of gas stream trailing the X-ray pulsar around the orbital phases 0.18-0.34. This would also be a possible explanation for the 
absence of second peak in the orbital distribution of N$_{H}$, though it is predicted both by \cite{pravdo2001} and \cite{leahy2008} models.

\section{Summary}

We have investigated the long term orbital variation of spectral parameters by exhaustive orbital phase resolved spectroscopic measurements of GX 301-2 with 
{\it MAXI}. The column density and flux of the Fe fluorescence line has a large value around the pre-periastron passage, suggesting the possible origin of X-ray flare due 
to enhanced mass accretion. A large column density and strong Fe emission line together with the presence of low energy excess in orbital phases around the 
X-ray flare, strongly favours a high density gas stream plus a stellar wind model for mode of accretion onto the neutron star in GX 301-2 \citep{leahy2008}. 
The orbital dependence of column density and equivalent width of Fe line presented in this work provides stronger constraints to the dynamical wind plus stream 
model of \cite{leahy2008}. The presence of very high equivalent width of Fe line for very low column density along line of sight at orbital phases 0.1-0.3 
places constrains on the direction of gas stream around GX 301-2 orbit. 
This has deep implications in understanding the interplay of accretion stream and stellar wind at different orbital phases. The relation between equivalent 
width of Fe line and column density of absorbing matter is a powerful tool which can be later used for probing the geometry and distribution of circumstellar 
matter around other wind-fed systems.

\vspace{1.5cm}

\textbf{ACKNOWLEDGMENT} \\
The authors are deeply grateful to Dr. T. Mihara for his helpful comments and suggestions. The authors also thank the anonymous referee for useful 
comments. This research has made use of {\it MAXI} data provided by RIKEN, JAXA and the MAXI team. This research has also made use of quick-look results of 
{\it RXTE-ASM} obtained through High Energy Astrophysics Science Archive Research Center Online Service (HEASARC), provided by the NASA/Goddard Space Flight 
Center.

\end{document}